\documentclass[11pt,a4paper]{article}
\usepackage{amsmath,amssymb}
\usepackage{epsfig,graphicx}

\title{\bf Axions and Cosmic Rays}

\author{D. Espriu$^a$$^b$\footnote{{\bf e-mail}: domenec.espriu@cern.ch}~ and
A. Renau$^{b}$\footnote{{\bf e-mail}: arencer@gmail.com}
\\
$^{a}$ \small{\em CERN, 1211 Geneva 23, Switzerland}
\\
$^{b}$\small{\em Departament d'Estructura i Constituents de la Mat\`eria and}\\
\small{\em Institut de Ci\`encies del Cosmos (ICCUB), Universitat de Barcelona,}\\
\small{\em Mart\'\i ~i Franqu\`es 1, 08028 Barcelona, Spain.}\\}

\date{\small{ To appear in Proceedings of the Quarks 2010 International Seminar,\\
 June 2010, Kolomna 
(Russia)}}

\begin{document}

\maketitle

\begin{abstract}
We investigate the propagation of a charged particle in a
spatially constant but time dependent pseudoscalar background.
Physically this pseudoscalar background could be  provided
by a relic axion density. The background leads to an explicit breaking of
Lorentz invariance; as a consequence processes such as $p\to p \gamma$ or $e\to e \gamma$ are possible within some
kinematical constraints. The phenomenon is described by the QED lagrangian extended with a Chern-Simons term that contains a 4-vector which characterizes the breaking of Lorentz invariance induced by the
time-dependent background. While the radiation induced (similar to the Cherenkov
effect) is too small to influence the propagation of cosmic rays in a significant way,
the hypothetical detection of the photons radiated  by
high energy cosmic rays via this mechanism would provide an indirect way of
verifying the cosmological relevance of axions. We discuss on the order of magnitude of the effect.
\end{abstract}

\vfill
\begin{flushright}%
ICCUB-10-062 \\
UB-ECM-PF-10-36
\end{flushright}


\newcommand{\be}{\begin{equation}}
\newcommand{\ee}{\end{equation}}
\newcommand{\nn}{\noindent}
\newcommand{\no}{\nonumber\\}
\newcommand{\ba}{\begin{eqnarray}}
\newcommand{\ea}{\end{eqnarray}}
\newcommand{\si}{\sin\theta}
\newcommand{\sii}{\sin^2\theta}
\newcommand{\co}{\cos\theta}
\newcommand{\coo}{\cos^2\theta}

\section{Axions}
Cold relic axions resulting from vacuum misalignment\cite{sikivie,raffelt}
in the early universe
is a popular and so far viable candidate to dark matter. If we assume that
cold axions are the only contributors to the matter density of the universe
apart from ordinary baryonic matter its density must be\cite{wmap}
\be
\rho\simeq 10^{-30}{\rm g} {\rm cm}^{-3}\simeq 10^{-46} {\rm GeV}^4.
\ee
Of course dark matter is not uniformly distributed, its distribution
traces that of visible matter (or rather the other way round). The
galactic halo of dark matter (assumed to consist of axions) would correspond to a
typical value for the density\cite{ggt}
\be
\rho_a\simeq 10^{-24}{\rm g} {\rm cm}^{-3}\simeq 10^{-40} {\rm GeV}^4
\ee
extending over a distance of 30  to 100 kpc in a galaxy such as
the Milky Way. Precise details of the density
profile are not so important at this point.
The axion background provides a very diffuse concentration of pseudoscalar particles
interacting very weakly with photons and therefore indirectly with
cosmic rays. What are the consequences
of this diffuse axion background on high-energy cosmic ray propagation?
Could this have an impact on cosmic ray propagation similar to the
GZK cutoff \cite{GZK}? This is the
question we would like to address here.

The fact that the axion is a pseudoscalar, being the pseudo Goldstone
boson of the broken Peccei-Quinn symmetry\cite{PQ}, is quite relevant.
Its coupling to photons will take place through the anomaly
term; hence the coefficient is easily calculable once the axion model is known
\be\label{term}
\Delta{\cal L}= g_{a\gamma\gamma} \frac{\alpha}{2\pi} \frac{a}{f_a}\tilde F F.
\ee
Two popular axion models are the DFSZ\cite{DFSZ} and
the KSVZ\cite{KSVZ} ones . In both models $g_{a\gamma\gamma}\simeq 1$. Here $a$ is
the axion field and $f_a$ is the axion decay constant. Further details are provided in section 3.

\section{Cosmic Rays}
Cosmic rays consist of particles (such as electrons, protons, helium and other nuclei) reaching the Earth from
outside. Primary cosmic rays are those produced at astrophysical sources (e.g. supernovae), while secondary cosmic
rays are particles produced by the interaction of primaries with interstellar gas.
In this work, the effect of
axions on the propagation of these cosmic rays will be studied. We will separately consider proton and
electron cosmic rays and ignore heavier nuclei because the effect on them will be far less important
as will become clear later (the axion-induced Bremsstrahlung depends on the mass of the charged
particle).


\subsection{Cosmic Ray Energy Spectrum}
We are interested in the number of protons in cosmic rays. Experimentally, one sees that the number of 
cosmic ray particles with a given energy depends on energy according to a power law
\be\label{flux} J(E)=
N_i E^{-\gamma_i},\ee where the spectral index $\gamma_i$ takes different values 
in different regions of the spectrum (see \cite{auger}).

For protons we have
\be
J_p(E)=\left\{\begin{array}{ll}
                     5.87 \cdot 10^{19} E^{-2.68} &  10^9\le E\le4\cdot10^{15} \\
                     6.57 \cdot 10^{28}E^{-3.26} & 4\cdot10^{15}\le E\le4\cdot10^{18} \\
                     2.23  \cdot 10^{16}E^{-2.59} & 4\cdot10^{18}\le E\le2.9\cdot10^{19} \\
                     4.22  \cdot 10^{49}E^{-4.3} &  E\ge2.9\cdot10^{19}
                   \end{array}
\right.
\label{cosmic1},
\ee
while for electrons the power law is\cite{electrons}
\be
J_e(E)=\left\{\begin{array}{cc}
                     5.87 \cdot 10^{17} E^{-2.68} & E\le5\cdot10^{10} \\
                     4.16 \cdot 10^{21} E^{-3.04} & E\ge5\cdot10^{10}
                   \end{array}
\right.
\label{cosmic2}
\ee
and the flux typically two orders of magnitude below that of protons, although it is more poorly known.
Our ignorance on electron cosmic rays is quite regrettable as it has
a substantial impact in our estimation of the radiation yield.

\begin{figure}[h]
\center
\includegraphics[scale=0.4]{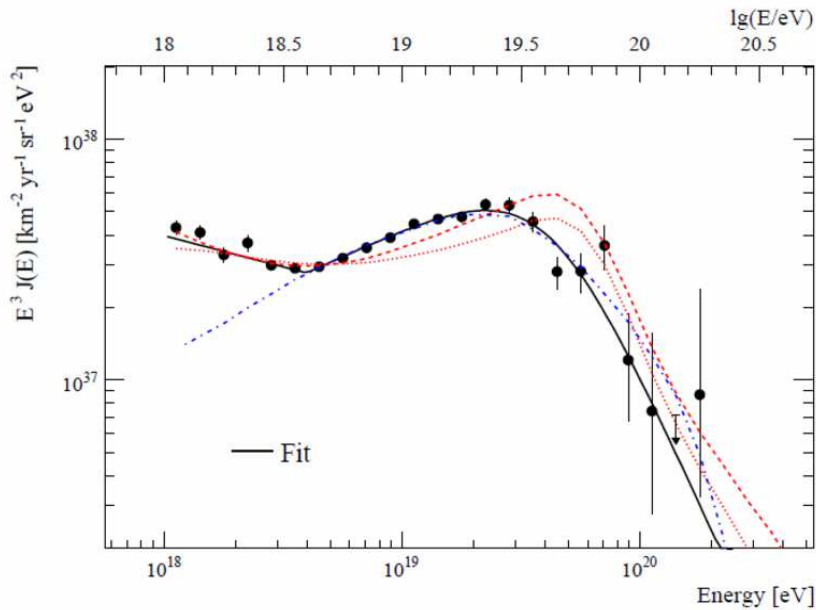}
\caption{Proton cosmic ray energy spectrum}
\label{spectrum}
\end{figure}

Note that the above ones are values measured locally in the inner solar system. It is known that
the intensity of cosmic rays increases with distance from the sun because the
modulation due to the solar wind makes more difficult for them to reach us, particularly so for electrons.
In addition, the hypothesis of homogeneity
and isotropy holds for proton cosmic rays, but not necessarily for electron
cosmic rays. Indeed because cosmic rays are deflected by magnetic fields they
follow a nearly random trajectory within
the Galaxy. We know that on average a hadronic
cosmic ray spends about $10^7$ years in the galaxy before escaping into intergalactic space.
This ensures the uniformity of the flux, at least for protons of galactic origin.
On the contrary, electron cosmic rays
travel for approximately 1 kpc on average before being slowed down.
However, because $l\sim  t^{1/2}$ for a random walk,
1 kpc corresponds to a typical age of an electron cosmic ray  $\sim 10^5$ yr\cite{kin}.
In addition, the lifetime of an electron cosmic ray depends
on the energy in the following way
\be
t(E)\simeq 5 \times 10^5 (\frac{1 \ {\rm TeV}}{E})\, {\rm yr}=\frac{T_0}E \label{ageee},
\ee
with $T_0 \simeq 2.4 \times 10^{40}$. 
To complicate matters further, it has been argued that the local interstellar flux of electrons is not
even representative of the Galaxy one and may reflect the electron debris from a nearby supernova
$\sim 10^4$ years ago\cite{age}.

\subsection{The GZK Cut-off}
The GZK (Greisen-Zatsepin-Kuzmin) limit\cite{GZK} states that the number of cosmic rays above a
certain energy threshold should be very small. Cosmic rays particles
interact with photons from the Cosmic Microwave Background (CMB) to produce pions
\be
\gamma_{\rm CMB}+p\longrightarrow p+\pi^0\quad \text{or}\quad \gamma_{\rm CMB}+p\longrightarrow n+\pi^+.
\ee
The energy threshold is about $10^{20}$ eV.
Because of the mean free path associated with these reactions, cosmic rays with energies above
the threshold and traveling over distances larger than 50 Mpc should not be observed on Earth.
This is the reason of the rapid fall off of the proton cosmic ray spectrum above $10^{20}$ eV as
there are very few nearby sources capable of providing such tremendous energies.

Note that the change in slope of the spectrum at around $10^{18}$ eV is believed to be due to the
appearance at that energy of extragalactic cosmic rays.

\section{Solving QED in a Cold Axion Background}
In this section we shall describe in great detail the theoretical tools needed to understand the
interactions between the highly energetic cosmic rays we have just described and the cold axion
background described in the first section.

The interaction of axions and photons is described by the following piece in the lagrangian
\be\label{prima}
\mathcal L_{a\gamma\gamma}=g_{a\gamma\gamma}\frac\alpha{2\pi}\frac a{f_a}F^{\mu\nu}\tilde F_{\mu\nu},
\ee
where
\be
\tilde F^{\mu\nu}=\frac12\varepsilon^{\mu\nu\alpha\beta}F_{\alpha\beta}
\ee
is the dual field strength tensor.

The axion field is originally misaligned and in the process of relaxing to
the equilibrium configuration coherent oscillations with ${\bf q}=0$ are produced,
provided that the reheating temperature after inflation is below the
Peccei-Quinn transition scale\cite{PQ}. In late
times the axion field evolves according to

\be\label{osc}
a(t)=a_0\cos(m_at),
\ee
where the amplitude $a_0$ is related to the initial misalignment angle. With this, \eqref{prima} becomes
\be
\mathcal L_{a\gamma\gamma}=g_{a\gamma\gamma}\frac\alpha{2\pi}\frac 1{f_a}a_0\cos(m_at)F^{\mu\nu}\tilde F_{\mu\nu}=
g_{a\gamma\gamma}\frac\alpha{\pi f_a}a_0\cos(m_at)\epsilon^{\mu\nu\alpha\beta}\partial_\mu A_\nu F_{\alpha\beta}.
\ee
Integrating by parts (dropping total derivatives) and taking into account that $\epsilon^{\mu\nu\alpha\beta}\partial_\mu F_{\alpha\beta}=0$, we get
\be\label{parts}
\mathcal L_{a\gamma\gamma}=g_{a\gamma\gamma}\frac{\alpha m_aa_0}{\pi f_a}\sin(m_at)\epsilon^{ijk}A_iF_{jk},
\ee
where Latin indices run over the spatial components only.

A cosmic ray particle (which travels at almost the speed of light) will see regions with quasi-constant values of the axion background, of a size depending on the axion mass, but always many orders of magnitude bigger than its wavelength. Thus, we can approximate the sine in \eqref{parts} by a constant ($\frac12$, for example). Then, it can be written as
\be\label{eq:Lagg}
\mathcal L_{a\gamma\gamma}=\frac14\eta_\mu A_\nu\tilde F^{\mu\nu},
\ee
where $\eta^\mu=(\eta,0,0,0)$ and $\displaystyle\eta=4g_{a\gamma\gamma}\frac{\alpha m_aa_0}{\pi f_a}$. The ``constant'' $\eta$ changes sign with a period $\displaystyle\sim 1/{m_a}$.

The oscillator has energy density $\displaystyle\rho_a=\frac12\dot a^2_{\rm max}=\frac12(m_aa_0)^2$, so $m_aa_0=\sqrt{2\rho_a}$.
Then, the constant $\eta$ is
\be
\eta=g_{a\gamma\gamma}\frac{4\alpha}\pi\frac{\sqrt{2\rho_a}}{f_a}\sim10^{-20}\,\rm eV,
\ee
for $\rho_a=10^{-4}\,\rm eV^4$ and $f_a=10^{7}\,\rm GeV=10^{16}\,\rm eV$.

The extra term in \eqref{eq:Lagg} corresponds to Maxwell-Chern-Simons Electrodynamics. Although in Maxwell-Chern-Simons Electrodynamics one can have in principle any four-vector $\eta^\mu$, the axion background provides a purely temporal vector. We shall assume  $\eta^\mu$ to be constant within a time interval $1/m_a$.  

\subsection{Euler-Lagrange Equations}
In the presence of an axion background the QED Lagrangian
is
\be\label{lagr}
\mathcal L=-\frac14F^{\mu\nu}F_{\mu\nu}+\bar\psi\left(i\not\!\partial-e\not\!\!A-m_e\right)\psi+\frac12m_\gamma^2 A_\mu A^\mu+\frac14\eta_\mu A_\nu\tilde F^{\mu\nu}.
\ee
Here also an  effective photon mass has been considered
(equivalent to a refractive index, see \cite{raffelt}). It is of order
\be
m_\gamma^2\simeq4\pi\alpha\frac{n_e}{m_e}.
\ee
The electron density in the Universe is expected to be at most $n_e\simeq10^{-7}\rm\,cm^{-3}\simeq10^{-21}\,\rm eV^3$.
This density corresponds to $m_\gamma\simeq10^{-15}\rm\,eV$, but the more conservative limit (compatible with \cite{pdg})
$m_\gamma=10^{-18}\,\rm eV$ will be used here.

The second term of \eqref{lagr} gives the kinetic and mass term for the fermions and also their interaction with photons. Dropping it, we get the Lagrangian for (free) photons in the axion background (see \cite{AGS} for further details):
\ba
\mathcal L&=&-\frac14F^{\mu\nu}F_{\mu\nu}+\frac12m_\gamma A_\mu A^\mu+\frac14\eta_\mu A_\nu\tilde F^{\mu\nu}\cr
&=&-\frac12\partial_\mu A_\nu(\partial^\mu A^\nu-\partial^\nu A^\mu)+\frac12m_\gamma^2A_\mu A^\mu+\frac14\epsilon^{\mu\nu\alpha\beta}\eta_\mu A_\nu\partial_\alpha A_\beta.
\ea
The Euler-Lagrange (E-L) equations are
\be
\partial_\sigma\frac{\partial\mathcal L}{\partial(\partial_\sigma A_\lambda)}-\frac{\partial\mathcal L}{\partial A_\lambda}=0,
\ee
\ba
\frac{\partial\mathcal L}{\partial A_\lambda}&=&\frac{\partial}{\partial A_\lambda}\left(\frac12m_\gamma^2g^{\mu\nu}A_\mu A_\nu+\frac14\epsilon^{\mu\nu\alpha\beta}\eta_\mu A_\nu\partial_\alpha A_\beta\right)\cr
&=&\frac12m^2_\gamma(g^{\lambda\nu}A_\nu+g^{\mu\lambda}A_\mu)+\frac14\epsilon^{\mu\lambda\alpha\beta}\eta_\mu\partial_\alpha A_\beta \cr
&=&m^2_\gamma A^\lambda+\frac14\epsilon^{\mu\lambda\alpha\beta}\eta_\mu\partial_\alpha A_\beta.
\ea
\ba
\partial_\sigma\frac{\partial\mathcal L}{\partial(\partial_\sigma A_\lambda)}&=&\partial_\sigma\frac{\partial}{\partial(\partial_\sigma A_\lambda)}\left[-\frac12\partial_\mu A_\nu g^{\alpha\mu}g^{\beta\nu}(\partial_\alpha A_\beta-\partial_\beta A_\alpha)+\frac14\epsilon^{\mu\nu\alpha\beta}\eta_\mu A_\nu\partial_\alpha A_\beta\right]\cr
&=&\partial_\sigma\left\{-\frac12\left[g^{\alpha\sigma}g^{\beta\lambda}(\partial_\alpha A_\beta-\partial_\beta A_\alpha)+\partial_\mu A_\nu(g^{\sigma\mu}g^{\lambda\nu}-g^{\lambda\mu}g^{\sigma\nu})\right]+\frac14\epsilon^{\mu\nu\sigma\lambda}\eta_\mu A_\nu\right\}\cr
&=&\partial_\sigma\left[-(\partial^\sigma A^\lambda-\partial^\lambda A^\sigma)+\frac14\epsilon^{\mu\nu\sigma\lambda}\eta_\mu A_\nu\right]\cr
&=&-\partial_\sigma\partial^\sigma A^\lambda+\partial^\lambda\partial_\sigma A^\sigma+\frac14\epsilon^{\mu\nu\sigma\lambda}\eta_\mu \partial_\sigma A_\nu.
\ea
Rearranging the indices, the equations are
\be
-\Box A^\lambda+\partial^\lambda\partial_\sigma A^\sigma-m^2A^\lambda-\frac12\epsilon^{\beta\lambda\mu\alpha}\eta_\mu\partial_\alpha A_\beta=0.
\ee
If we choose the Lorenz gauge $\partial_\alpha A^\alpha=0$ the second term vanishes. The equations can also be written as
\be
-g^{\beta\lambda}\Box A_\beta-g^{\lambda\beta}m^2_\gamma A_\beta-\frac12\epsilon^{\beta\lambda\mu\alpha}\eta_\mu\partial_\alpha A_\beta=0.
\ee
We are interested in writing these equations in momentum space. To this end, define the Fourier transform of the field:
\be
A_\mu(x)=\int\frac{d^4k}{(2\pi)^4}e^{-ikx}\tilde A_\mu(k).
\ee
The relevant derivatives are
\be
\partial_\alpha A_\beta=\int\frac{d^4k}{(2\pi)^4}(-ik_\alpha)e^{-ikx}\tilde A_\beta(k)
\ee
and
\be
\Box A_\beta=\int\frac{d^4k}{(2\pi)^4}(-k^2)e^{-ikx}\tilde A_\beta(k).
\ee
The E-L equations are then
\be
\int\frac{d^4k}{(2\pi)^4}\left[g^{\beta\lambda}(k^2-m^2_\gamma)+\frac i2\epsilon^{\beta\lambda\mu\alpha}\eta_\mu k_\alpha\right] e^{-ikx}\tilde A_\beta(k)=0.
\ee
Therefore,
\be\label{eqB}
\left[g^{\beta\lambda}(k^2-m^2_\gamma)+\frac i2\epsilon^{\beta\lambda\mu\alpha}\eta_\mu k_\alpha\right]\tilde A_\beta(k)=0,
\ee
or
\be\label{eqA}
K^{\mu\nu}\tilde A_\nu(k)=0,\qquad K^{\mu\nu}=g^{\mu\nu}(k^2-m^2_\gamma)+\frac i2\epsilon^{\mu\nu\alpha\beta}\eta_\alpha k_\beta.
\ee
\subsection{Polarization Vectors and Dispersion Relation}
We now define
\be
S^\nu_{\,\lambda}=\epsilon^{\mu\nu\alpha\beta}\eta_\alpha k_\beta\epsilon_{\mu\lambda\rho\sigma}\eta^\rho k^\sigma.
\ee
This can be put in a more convenient form using the contraction of two Levi-Civita symbols
$\epsilon_{\mu\lambda\rho\sigma}\epsilon^{\mu\nu\alpha\beta}=-3!\delta_{[\lambda}^\nu\delta_\rho^\alpha\delta_{\sigma]}^\beta$ (the minus sign is there because in Minkowski space $\epsilon_{0123}=-\epsilon^{0123}$):
\be\label{S}
S^{\mu\nu}=\left[\left(\eta\cdot k\right)^2-\eta^2 k^2\right]g^{\mu\nu}
- \left(\eta\cdot k\right)\left(\eta^\mu k^\nu + k^\mu\eta^\nu \right)
+ k^2\eta^\mu\eta^\nu + \eta^2 k^\mu k^\nu.
\ee
It satisfies
\be
S^\mu_{\,\nu}\eta^\nu=S^\mu_{\,\nu}k^\nu=0,\quad S=S^\mu_\mu=2\left[\left(\eta\cdot k\right)^2-\eta^2 k^2\right],\quad
S^{\mu\nu}S_{\nu\lambda}=\frac S2S^\mu_{\,\lambda}.
\ee
If $\eta^\mu=(\eta,0,0,0)$ we have $S=2\eta^2\vec k^2>0$.
Now we introduce two projectors:
\be
P^{\mu\nu}_\pm=\frac{S^{\mu\nu}}S\mp\frac i{\sqrt{2S}}\epsilon^{\mu\nu\alpha\beta}\eta_\alpha k_\beta.
\ee
These projectors have the following properties:
\ba\label{properties}
P^{\mu\nu}_{\pm}\eta_{\nu}=P^{\mu\nu}_{\pm}k_{\nu}=0,\quad
g_{\mu\nu}P^{\mu\nu}_{\pm}=1,\quad  (P_\pm^{\mu\nu})^*=P_\mp^{\mu\nu}=P_\pm^{\nu\mu} ,\cr
P^{\mu\lambda}_{\pm}P_{\pm\lambda\nu}=P^{\mu}_{\pm\nu},\quad P^{\mu\lambda}_{\pm}\,P_{\mp\lambda\nu} = 0,\quad P^{\mu\nu}_{+} + P^{\mu\nu}_{-}= \frac{2}{S}S^{\mu\nu}.
\ea
With these projectors, we can build  a pair of polarization vectors to solve \eqref{eqA}. We start from a space-like unit vector, for example $\epsilon=(0,1,1,1)/\sqrt3$. Then, we project it:
\be
\tilde\varepsilon^\mu=P_\pm^{\mu\nu}\epsilon_\nu.
\ee
In order to get a normalized vector, we need
\ba
(\tilde\varepsilon^\mu_\pm)^*\tilde\varepsilon_{\pm\mu}&=&P_\pm^{\nu\mu}\epsilon_\nu P_{\pm\mu\lambda}\epsilon^\lambda=P^\nu_{\pm\lambda}\epsilon_\nu\epsilon^\lambda=\frac{S^{\nu\lambda}\epsilon_\nu\epsilon_\lambda}S\cr
&=&\frac{S/2\epsilon^\mu\epsilon_\mu+\eta^2(\epsilon\cdot k)^2}S
=-\frac12+\frac{(\epsilon\cdot k)^2}{2\vec k^2}
\ea(this is of course negative because $\epsilon$ is space-like).
Then, the polarization vectors are
\be
\varepsilon^\mu_\pm=\frac{\tilde\varepsilon^\mu_\pm}{\sqrt{-\tilde\varepsilon^\nu_\pm\tilde\varepsilon^*_{\pm\nu}}}=
\left[\frac{\vec k^2-(\epsilon\cdot k)^2}{2\vec k^2}\right]^{-1/2}P_\pm^{\mu\nu}\epsilon_\nu.
\ee
These polarization vectors satisfy
\be\label{orto}
g_{\mu\nu}\varepsilon_\pm^{\mu*}\varepsilon_\pm^\nu=-1,\quad g_{\mu\nu}\varepsilon_\pm^{\mu*}\varepsilon_\mp^\nu=0
\ee
and
\be\label{closure}
\varepsilon_\pm^{\mu*}\varepsilon_\pm^\nu+\varepsilon_\pm^\mu\varepsilon_\pm^{\nu*}=-\frac2SS^{\mu\nu}=-\frac{S^{\mu\nu}}{\eta^2\vec k^2}
\ee
With the aid of the projectors, we can write the tensor in \eqref{eqA} as
\be
K^{\mu\nu}=g^{\mu\nu}(k^2-m_\gamma^2)+\sqrt{\frac S2}\left(P_-^{\mu\nu}-P_+^{\mu\nu}\right).
\ee
Then we have for $k=(\omega_\pm,\vec k)$
\be
K^\mu_\nu\varepsilon^\nu_\pm=\left[(k^2-m_\gamma^2)\mp\sqrt{\frac S2}\right]\varepsilon^\nu_\pm=
\left(k^2-m_\gamma^2\mp\eta|\vec k|\right)\varepsilon^\mu_\pm=\left(\omega_\pm^2-\vec k^2-m_\gamma^2\mp\eta|\vec k|\right)\varepsilon^\mu_\pm.
\ee
Therefore, $\tilde A^\mu=\varepsilon_\pm^\mu$ is a solution of \eqref{eqA} iff
\be
\omega_\pm(\vec k)=\sqrt{m_\gamma^2\pm\eta|\vec k|+\vec k^2}.
\ee
This is the new dispersion relation of photons in the cold axion background in the approximation where
$\eta$ is assumed to be piecewise constant.

\section{The Process $p\longrightarrow p\,\gamma$}

\subsection{Kinematic Constraints}
We now consider $p(p)\longrightarrow p(q)\gamma(k)$, or $e(p)\longrightarrow e(q)\gamma(k)$. This process is forbidden in normal QED due to the conservation of energy. It is, however, possible in this background (the cold axion background even allows the process $\gamma\rightarrow e^+e^-$, see \cite{AEGS}). Momentum conservation means $\vec q=\vec p-\vec k$. Calling $m$ the mass of the charged particle (proton or electron), conservation of energy leads to
\be
\begin{array}{c}\label{cons}
  E(q)+\omega(k)=E(p),
  \sqrt{m^2+(\vec p-\vec k)^2}+\sqrt{m_\gamma^2\pm\eta|\vec k|+\vec k^2}=\sqrt{m^2+\vec p^2}, \\
  \sqrt{E^2+k^2-2pk\co}+\sqrt{m_\gamma^2\pm\eta k+k^2}-E=0
\end{array}
\ee
In the last line, a lighter notation has been adopted:
\be
E=E(p)=\sqrt{m^2+\vec p^2},\quad p=|\vec p|,\quad k=|\vec k|,\quad\vec p\cdot\vec k=pk\co.
\ee
As will be seen, if $\eta$ is positive (negative) the process is only possible for negative (positive) polarization. Therefore, $\pm\eta=-|\eta|$ in these cases. To take into account both of them, we will use the minus sign and write $\eta$ instead of $|\eta|$.

Squaring twice yields
\be
(4E^2-4p^2\coo+4p\eta\co-\eta^2)k^2-2(2E^2\eta+2m^2_\gamma p\co-m_\gamma^2\eta)k+(4E^2m_\gamma^2-m_\gamma^4)=0.
\ee
Neglecting $m_\gamma,\eta$ in front of $m,E$ this is:
\be
(E^2-p^2\coo+p\eta\co)k^2-(E^2\eta+m^2_\gamma p\co)k+E^2m_\gamma^2=0.
\ee
This equation has two solutions
\be\label{kpm}
k_\pm=\frac{E^2\eta+pm_\gamma^2\co\pm E\sqrt{E^2\eta^2-4E^2m_\gamma^2+4p^2m_\gamma^2\coo-2pm_\gamma^2\eta\co}}{2(E^2-p^2\coo+p\eta\co)}.
\ee
These solutions only make sense if the discriminant $\Delta$ is positive. With the approximation $\co\simeq1-\frac12\sii$, the condition $\Delta\ge0$ is
\be
\sii\le\frac{\left[p^2\eta^2-2pm^2_\gamma\eta+m^2(\eta^2-4m_\gamma^2)\right]}{4p^2m_\gamma^2(1-\frac\eta{4p})}.
\ee
Which can be rewritten as
\be\label{angle}
\sii\le\frac{\eta^2}{4p^2m_\gamma^2}\frac1{1-\frac\eta{4p}}(p-p_+)(p-p_-),
\ee
where
\be
p_\pm=\frac{m_\gamma^2}\eta\pm\frac{2mm_\gamma}\eta\sqrt{1-\frac{\eta^2}{4m_\gamma^2}}
\simeq\pm\frac{2mm_\gamma}\eta\sqrt{1-\frac{\eta^2}{4m_\gamma^2}}.
\ee
It is clear that $p_+>0$ and $p_-<0$. For $\sii$ to be positive we need
\be
p>p_+=p_{th}=\frac{2mm_\gamma}\eta\sqrt{1-\frac{\eta^2}{4m_\gamma^2}}.
\ee
This is the threshold below which the process cannot take place kinematically. The energy threshold \big($E_{th}^2=m^2+p_{th}^2$\big) is:
\be
E_{th}=\frac{2mm_\gamma}\eta.
\ee
When $\eta\rightarrow0$, the threshold goes to infinity (as is expected: the process cannot happen if $\eta$ vanishes).

There is another relevant scale in the problem: $m^2/\eta$. It is many orders of magnitude above the GZK cut-off. Therefore, we will always assume the limit $p\ll m^2/\eta$.
The maximum angle of emission for a given momentum is given by \eqref{angle}:
\be
\sii_{\rm max}(p)=\frac{\eta^2}{4p^2m_\gamma^2}\frac1{1-\frac\eta{4p}}(p-p_+)(p-p_-).
\ee
Its greatest value is obtained when $p$ is large ($p\gg p_{th}$):
\be
\sii_{\rm max}=\frac{\eta^2}{4m_\gamma^2}.
\ee
Since this is a small number, photons are emitted in a narrow cone $\theta_{\rm max}=\frac\eta{2m_\gamma}$. This justifies the approximation made for $\co$.

At $\theta_{\rm max}(p)$, the square root in \eqref{kpm} vanishes and
\be
k_+[\theta_{\rm max}(p)]=k_-[\theta_{\rm max}(p)]=\xrightarrow{p_{th}\ll p\ll m^2/\eta}\frac{2m_\gamma^2}\eta.
\ee
The minimum value for the angle is $\theta=0$:
\be
k_\pm(0)\simeq\frac{E^2\eta+pm_\gamma^2\pm\left(E^2\eta-pm_\gamma^2-2\frac{m^2m_\gamma^2}\eta\right)}{2(m^2+p\eta)}.
\ee
This gives the maximum and minimum values of the photon momentum. In the limit $p_{th}\ll p\ll m^2/\eta$ they are:
\be\label{kmax}
k_{\rm max}=k_+(0)=\frac{\eta E^2}{m^2}.
\ee
\be\label{kmin}
k_{\rm min}=k_-(0)=\frac{m_\gamma^2}\eta.
\ee
These two values coincide at the energy threshold.

Here we can see that the process is possible for negative (positive) polarization only if $\eta>0$ ($\eta<0$). 
Otherwise, the modulus of the photon momentum would be negative.

Note that the incoming cosmic ray wavelength fits perfectly within the $1/m_a$ size, so it indeed sees an
almost perfectly constant $\eta$. Whether $\eta$ is positive or negative there is {\em always} a state with slightly
less energy to which decay and lose part of its energy (of ${\cal O}(\eta)$) emitting a soft photon.
So even if the process
is a rare one it does not average to zero. An exact analysis will we presented elsewhere.

\begin{figure}[h]
\center
\includegraphics[scale=0.9]{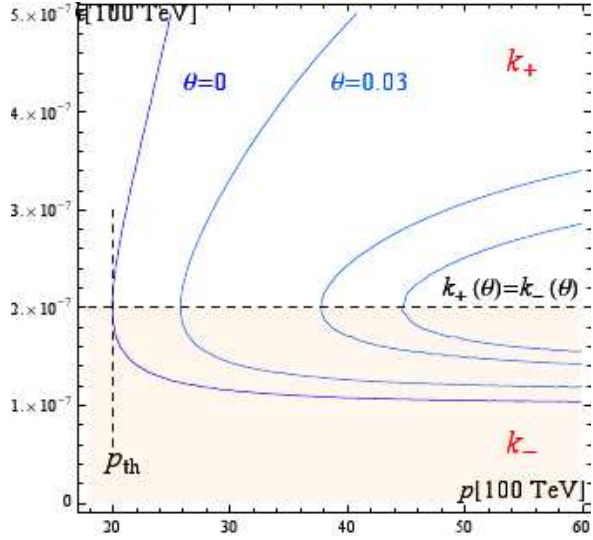}
\caption{The solution $k_\pm$ of the energy conservation equation \eqref{cons}. It can be seen that \eqref{kmax} and \eqref{kmin} are indeed the maximum and minimum values.}
\end{figure}

\subsection{Amplitude}
The next thing we need is to compute the matrix element for the process. Using the standard Feynman rules we get
\be
i\mathcal M=\bar u(q)ie\gamma^\mu u(p)\varepsilon^*_\mu(k).
\ee
Its square is
\be
|\mathcal M|^2=\bar u(q)ie\gamma^\mu u(p)\varepsilon^*_\mu(k)[\bar u(q)ie\gamma^\nu u(p)\varepsilon^*_\nu(k)]^*=e^2\varepsilon^*_\mu(k)\varepsilon_\nu(k)tr\left[u(q)\bar u(q)\gamma^\mu u(p)\bar u (p)\gamma^\nu\right].
\ee
We now must sum and average over initial and final proton helicities, respectively. We do not average over photon polarizations because the process is possible only for one polarization. Performing the trace:
\ba
\overline{|\mathcal M|^2}
&=&\frac12e^2\varepsilon^*_\mu(k)\varepsilon_\nu(k)tr[(\not\!q+m)\gamma^\mu(\not\!p+m)\gamma^\nu]\cr
&=&\frac12e^2\varepsilon^*_\mu(k)\varepsilon_\nu(k)tr[\not\!q\gamma^\mu\!\!\!\not\!p\gamma^\nu+m^2\gamma^\mu\gamma^\nu]\cr
&=&\frac12e^2\varepsilon^*_\mu(k)\varepsilon_\nu(k)[q^\mu p^\nu-q^\alpha p_\alpha g^{\mu\nu}+q^\nu p^\mu+m^2g^{\mu\nu}].
\ea
Using 4-momentum conservation, \eqref{orto} and the fact that $p^\alpha p_\alpha=m^2$, we get
\be
\overline{|\mathcal M|^2}=2e^2[-p^\alpha k_\alpha+2\varepsilon^*_\mu\varepsilon_\nu p^\mu p^\nu]=
2e^2\left[-p^\alpha k_\alpha+\left(\varepsilon^*_\mu\varepsilon_\nu+\varepsilon_\mu\varepsilon^*_\nu\right)p^\mu p^\nu\right].
\ee
Now we use \eqref{closure} to get
\be
\left(\varepsilon^*_\mu\varepsilon_\nu+\varepsilon_\mu\varepsilon^*_\nu\right)p^\mu p^\nu=-\frac{S^{\mu\nu}p_\mu p_\nu}{\eta^2k^2}=p^2\sii.
\ee
The averaged square amplitude is then
\be
\overline{|\mathcal M|^2}=2e^2\left(-p^\alpha k_\alpha+p^2\sii\right).
\ee
The first term is positive:
\be
-p^\alpha k_\alpha=-E\omega+pk\co=-E\omega-pk\frac{m_\gamma^2-\eta k-2E\omega}{2pk}=\frac12{\eta}(k-\frac{m_\gamma^2}\eta)=\frac12(k-k_{\rm min})>0,
\ee
so $\overline{|\mathcal M|^2}$ is clearly positive.
\subsection{Differential Decay Width}
The differential decay width is
\be
d\Gamma=(2\pi)^4\delta^{(4)}(q+k-p)\frac1{2E}\overline{|\mathcal M|^2}dQ,
\ee
where the phase space element is
\be
dQ=\frac{d^3q}{(2\pi)^32E(q)}\frac{d^3k}{(2\pi)^32\omega(k)}.
\ee
We can use $\delta^{(3)}(\vec q+\vec k-\vec p)$ to eliminate $d^3q$. The remaining $\delta$ is the conservation of energy. Therefore, $E(q)=E-\omega$. Next we use a property of the Dirac delta function
\be
\delta[f(x)]=\sum_i\frac{\delta(x-x_i)}{|f'(x_i)|},
\ee
where $x_i$ are the zeros of the function. In our case, we consider $E(q)$ a function of $\co$:
\ba
\delta[E(q)+\omega-E]&=&\delta\left(\sqrt{E^2+k^2-2pk\co}+\omega-E\right)\cr
&=&\left|\frac{-2pk}{2\sqrt{E^2+k^2-2pk\co}}\right|^{-1}
\delta\left(\co-\frac{m_\gamma^2-\eta k-2E\omega}{-2pk}\right).
\ea
Next we write $d^3k=k^2dkd(\co)d\varphi$, integrate the $\varphi$ angle (factor of $2\pi$) and use the delta to eliminate $d(\co)$. This fixes the value of $\co$:
\be\label{cosine}
\co=\frac{m_\gamma^2-\eta k-2E\omega}{-2pk}.
\ee
Finally, the differential decay width is
\be
d\Gamma=\frac\alpha2\frac k{Ep\omega}(-p^\alpha k_\alpha+p^2\sii)dk,
\ee
where $\alpha=e^2/4\pi$ and $\si$ is given by \eqref{cosine}.
This decay width can be written more conveniently for future computations:
\be
\frac{d\Gamma}{dk}=\frac\alpha8\frac1{k\omega}\left[A(k)+B(k)E^{-1}+C(k)E^{-2}\right]
\theta(\frac{E^2\eta}{m^2}-k),
\ee
with
\be
A(k)=4(\eta k - m_\gamma^2),\quad
B(k)=4\omega(m_\gamma^2-\eta k),\quad
C(k)=-2m_\gamma^2k^2+2\eta k^3-m^4_\gamma-\eta^2k^2+2m_\gamma^2\eta k.
\ee

\subsection{Effects on cosmic rays}

We now want to compute the energy loss of protons in this background
\be
\frac{dE}{dx}=\frac{dt}{dx}\frac{dE}{dt}=\frac1v\left(-\int\omega d\Gamma\right).
\ee
Using the previous results and $v=p/E$, the energy loss is (with the integration limits given by \eqref{kmax} and \eqref{kmin})
\ba
\frac{dE}{dx}&=&-\frac\alpha2\frac1{p^2}\int_{k_{\rm min}}^{k_{\rm max}} kdk\left[\frac12(\eta k-m_\gamma^2)+p^2(1-\coo)\right]\cr
&=&-\frac\alpha{8p^2}\int_{k_{\rm min}}^{k_{\rm max}}dk\Bigg[2\eta k^2-(4m^2+2m_\gamma^2+\eta^2)k+2\eta(2E^2+m_\gamma^2)\cr
&&-m_\gamma^2(4E^2+m_\gamma^2)\frac1k+4E\eta\sqrt{m_\gamma^2-\eta k+k^2}+4Em_\gamma^2\frac{\sqrt{m_\gamma^2-\eta k+k^2}}k\Bigg]\cr
&=&-\frac\alpha{8p^2}\Bigg[\frac23\eta\left(\frac{\eta^3E^6}{m^6}-\frac{m_\gamma^6}{\eta^3}\right)-
\frac12(4m^2+2m_\gamma^2+\eta^2)\left(\frac{\eta^2E^4}{m^4}-\frac{m_\gamma^4}{\eta^2}\right)\cr
&&+2\eta(2E^2+m_\gamma^2)\left(\frac{\eta E^2}{m^2}-\frac{m_\gamma^2}\eta\right)-m_\gamma^2(4E^2+m_\gamma^2)\ln\left(\frac{\eta^2E^2}{m^2m_\gamma^2}\right)+...\Bigg].
\ea
The leading term is
\be
\frac{dE}{dx}=-\frac\alpha{8p^2}\frac{2\eta^2E^4}{m^2}=-\frac{\alpha\eta^2E^2}{4m^2v^2}\simeq-\frac{\alpha\eta^2E^2}{4m^2}.
\ee
The energy as a function of the traveled distance is then
\be
E(x)=\frac{E(0)}{1+\frac{\alpha\eta^2}{4m^2}E(0)x}.
\ee
The fractional energy loss for a cosmic ray with initial energy E(0) traveling a distance $x$ is
\be
\frac{E(0)-E(x)}{E(0)}=\frac{\frac{\alpha\eta^2}{4m^2}E(0)x}{1+\frac{\alpha\eta^2}{4m^2}E(0)x}
\ee
This loss is more important the more energetic the cosmic ray is. However, $\frac{\alpha\eta^2}{4m^2}$
is a very small number. If we take $E(0)=10^{20}$ eV (the energy of the most energetic cosmic rays)
and $x=10^{26}$ cm (about the distance to Andromeda, the nearest galaxy, therefore
larger than the galactic halo)
the energy loss is smaller than  $1$ eV. For less energetic cosmic rays, the effect is even weaker.

As we have seen, the effect of the axion background on cosmic rays is quite negligible. However,
the emitted photons may be detectable. Using $m_\gamma= 10^{-18}$ eV and $\eta= 10^{-20}$ eV as
indicative values and having in mind the GZK cut-off for protons (and a similar one for electrons\footnote{It is very doubtful that electrons could be accelerated to such energies but it is irrelevant anyway for the present discussion as the intensity is extremely small at these energies})
the emitted photon momenta fall in the range
\be
10^{-16} {\rm ~eV} 
<  k  <
100 {\rm ~eV}
\ee for primary protons and
\be
10^{-16} {\rm ~eV}  < k  < 400 {\rm ~MeV}
\ee
for primary electrons.

The number of cosmic rays with a given energy crossing a surface element per unit time is
\be
d^3N=J(E)dEdSdt_0,
\ee
where $J(E)$ is the cosmic ray flux. These cosmic rays will radiate at a time $t$. The number of
photons is given by
\be
d^5 N_\gamma = d^3 N \frac{d\Gamma(E,k)}{dk} dk dt= J(E) \frac{d\Gamma(E,k)}{dk} dE dk dt_0 dS dt.
\ee
Assuming that the cosmic ray flux does not depend on time, we integrate over $t_0$ obtaining a
factor $t(E)$: the age of the average cosmic ray with energy $E$. Since we do not care
about the energy of the primary cosmic ray (only that of the photon matters),
we integrate also over $E$, starting from $E_{\rm min}(k)$, the minimum energy that the
cosmic ray can have in order to produce a photon with momentum k, given by \eqref{kmax}. Therefore,
the flux of photons is
\be
\frac{d^3 N_\gamma}{dk dS dt}=\int_{E_{\rm min}(k)}^\infty dE \; t(E) J(E)\frac{d\Gamma(E,k)}{dk}~,\qquad
E_{th}=2\frac{mm_\gamma}{\eta}. \label{yield}
\ee
Next we assume that $t(E)$ is approximately constant and take $t(E)\approx T_p=10^7$ yr for protons
and $t(E)\approx T_e=5\cdot10^5$ yr for electrons. We know that this last approximation is not
correct as $t(E)\sim 1/E$ but at this point we are just interested in getting an
order of magnitude estimate of the effect.

The photon energy flux is obtained by multiplying the photon flux \eqref{yield} by the
energy of a photon with momentum $k$:
\ba
I(k)&=&\omega(k)\int_{E_{min}(k)>E_{th}}^\infty dE\ t(E) J(E)\frac{d\Gamma}{dk}\\
&\approx&\frac{\alpha T}{8k}\int_{E_{min}(k)}^\infty dE\ N_i
\left[ A(k)E^{-\gamma_i}
+B(k)E^{-(\gamma_i+1)}
+C(k)E^{-(\gamma_i+2)}\right],
\ea
where $E_{min}(k)=m\sqrt{\frac k\eta }$, see \eqref{kmax}. Numerically, the only relevant term in the decay rate is $4\eta k$, from $A(k)$.
The integral can then be approximated by
\be
I(k)\simeq \frac{\alpha \eta T}{2} \frac{J\left[E_{min}(k)\right] E_{min}(k)}{\gamma_{min}-1}\propto k^{-\frac{\gamma-1}2}.\label{approx}
\ee
The value $\gamma_{min}$ is to be read from (\ref{cosmic1}) or (\ref{cosmic2}) depending
on the range where $E_{min}(k)$ falls.

Substituting the numerical values we obtain the following approximate expressions
for $I_p(k)$ and $I_e(k)$
\be
I_p(k)=6  \times \left( \frac{T_p}{10^7 \ {\rm yr}}\right) \left(\frac{\eta}{10^{-20}~{\rm eV}}\right)^{1.84}
 \left(\frac{k}{10^{-7}\ {\rm eV}}\right)^{-0.84}
{\rm ~m}^{-2}\, {\rm s}^{-1}\, {\rm sr}^{-1}.
\ee
\be
I_e(k)=200 \times \left( \frac{T_e}{5\times10^5 \ {\rm yr}}\right) \left(\frac{\eta}{10^{-20}~{\rm eV}}\right)^{2.02}
 \left(\frac{k}{10^{-7}\ {\rm eV}}\right)^{-1.02}
{\rm ~m}^{-2}\, {\rm s}^{-1}\, {\rm sr}^{-1}.
\ee
As mentioned above these expressions are only indicative and assume constant average values
for the age of a cosmic ray (either proton or electron). For a more detailed discussion we encourage
the reader to examine our recent paper \cite{emr}. From this latter work we include
the following figure describing the radiation yield

\begin{figure}[h]
\center
\includegraphics[scale=1.4]{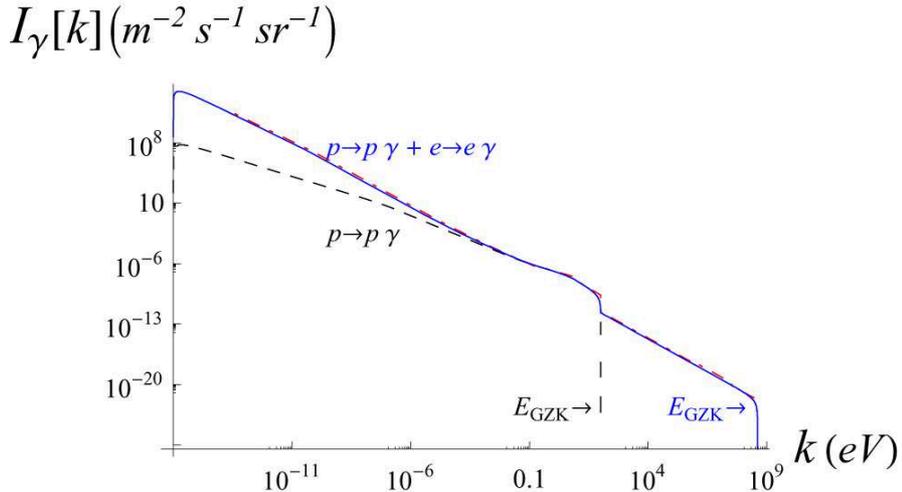}
\caption{Radiation yield using the exact formulae and a more appropriate parametrization of the
electron cosmic ray average lifetime as a function of the energy. From \cite{emr}. Note that electrons 
in general dominate the effect at low energies.}
\end{figure}

\section{Conclusions and Outlook}

In this work, the effect on charged particles of a mildly (compared to the
particle momentum) time dependent pseudoscalar background has been investigated.
We have been interested both in proton and electron cosmic rays.

This effect is calculable because the axion background
induces a modification of QED that is exactly solvable.
This modification has some interesting features, such as the
possibility of the photon emission process $p\rightarrow p\,\gamma$
and $e\rightarrow e\, \gamma$ (which we have termed as axion-induced
Bremsstrahlung processes). Kinematical constraints on the process have been
reviewed, in particular it is seen that it is only possible for proton
energies higher than a certain threshold. The energy loss of protons in such a
background has been computed. For protons that survive the GZK cutoff this loss is
totally negligible.

However, the radiated photons could still be detected. Their flux and
energy spectrum have been computed in some detail. Since the energy threshold depends
on the mass of the charged particle, it is lower for lighter particles. Also,
the energy loss is proportional to the mass squared of the charged particle, so the effect is
more important for electrons. The value of $k_{\rm min}$ does not depend on the charged particle mass,
so the radiated spectrum is no very different for electrons or protons (however the average
lifetime of electron and protons cosmic rays is quite different and this has an observable effect
on the power spectrum of the radiation).

We refer the interested reader
to \cite{emr} for a more comprehensive description of this phenomenon and on the
possibility of this diffuse radiation being measured.
We summarize however the main conclusions below.

The dominant contribution to the radiation
yield via this mechanism comes from electron (and positron) cosmic rays.
If one assumes that the power spectrum of the cosmic
rays is characterized by an exponent $\gamma$ then the produced radiation has an spectrum
$k^{-\frac{\gamma-1}{2}}$ for proton primaries, which becomes $k^{-\frac{\gamma}{2}}$ for electron primaries.
The dependence on the key parameter $\eta\sim \frac{\sqrt\rho_*}{f_a}$
comes with the exponent $\eta^{\frac{1+\gamma}{2}}$ and $\eta^{\frac{2+\gamma}{2}}$ for protons and electrons, respectively.
However for the regions where the radiation yield is largest electrons amply dominate. We have assumed that
the flux of electron cosmic rays
is uniform throughout the Galaxy and thus identical to the one observed in our neighbourhood, but relaxing
this hypothesis could provide an enhancement of the effect by a relatively large factor.
The effect for the lowest wavelengths where the atmosphere is transparent and for values of $\eta$ corresponding
to the current experimental limit is of ${\cal O}(10^{-1})$ mJy. This is at the limit of sensitivity
of antenna arrays that are already currently being deployed and thus a possibility worth exploring.

In the case of radiation originating from our galaxy the main unknown in the present
discussion is whether the flux of electron cosmic rays measured in our neighbourhood is
representative of the Galaxy or not. Since it is possible to relate this
flux to the galactic synchrotron radiation one could deduce the former from measuring the
latter. It appears\cite{synchro} that either the total number of electron cosmic rays is
substantially larger than the one measured in the solar system, or the galactic magnetic
fields have to stronger than expected. This issue remains to be further quantified.
No attempt has been made to quantify the signal from possible extragalactic sources either.

One should note that the effect discussed here
is a collective one. This is at variance with the GZK effect alluded in the first section
- the CMB radiation is not a coherent one
over large scales. For instance, no similar effect exists for hot axions.
A second observation is that some of the scales that play a role in the present discussion
are somewhat non-intuitive (for instance the 'cross-over' scale $m_p^2/\eta$ or the
threshold scale $m_\gamma m_p/\eta$). This is due to
the non Lorentz-invariant nature of this effect. Finally, it may look surprising at first that
an effect that has such a low probability may give a small but not ridiculously small contribution.
The reason why this happens
is that the number of cosmic rays is huge. It is known that they contribute to the energy density
of the Galaxy by an amount similar to the Galaxy's magnetic field\cite{adr}.

There are several aspects of the present analysis that could be improved to make it more precise, 
particularly a piecewise constant oscillating axion background, or one with a serrated time profile for that
matter, could be solved easily without having to appeal to special functions (the sinus profile
involves Mathieu functions). This will be presented elsewhere but the present analysis suffices
to indicate the order of magnitude of the effect.

We hope that the present mechanism help to assess
the presumed relevance of cold axions as a dark matter candidate.

\section*{Acknowledgements}

We thank our collaborator F. Mescia who participated in the research reported here. 
We acknowledge the financial support from projects FPA2007-66665,
2009SGR502, Consolider CPAN CSD2007-00042 and FLAVIANET. We thank A. De R\'ujula for
some comments on cosmic rays and, particularly, J.M. Paredes and P. Planesas
for discussions concerning the Galaxy synchrotron radiation.
D.E. would like to thank the organizers of the Quarks 2010 workshop in Kolomna for the
warm hospitality extended to him.


\end{document}